\documentclass[traditabstract]{aa} 

\usepackage{multirow}
\usepackage{array}
\usepackage{graphicx}
\usepackage{natbib}
\usepackage{lscape}
\usepackage{rotating}
\newcommand{\hei}{He\,{\sc i}}
\newcommand{\heii}{He\,{\sc ii}}
\newcommand{\ciii}{C\,{\sc iii}}
\newcommand{\niii}{N\,{\sc iii}}

\newcommand{\ha}{H$\alpha$}

\begin{document}

\title{Photometric identification of the periods of the first candidate extra-Galactic magnetic massive stars}

\author{Ya\"el Naz\'e\inst{1}\fnmsep\thanks{FNRS Research Associate}, Nolan R. Walborn\inst{3}, Nidia Morrell\inst{2}, Gregg A. Wade\inst{4}, Micha{\l} K. Szyma{\'n}ski\inst{5}}

\institute{Groupe d'Astrophysique des Hautes Energies, Institut d'Astrophysique et de G\'eophysique, Department AGO, Universit\'e de Li\`ege, 17, All\'ee du 6 Ao\^ut, B5c, B-4000 Sart Tilman, Belgium; \email{naze@astro.ulg.ac.be}
\and
Las Campanas Observatory, Carnegie Observatories, Casilla 601, La Serena, Chile
\and
Space Telescope Science Institute\thanks{Operated by AURA, Inc., under NASA contract NAS5-26555.}, 3700 San Martin Drive, Baltimore, MD 21218, USA
\and
Department of Physics, Royal Military College of Canada, PO Box 17000, Station Forces, Kingston, ON K7K 4B4, Canada
\and
Warsaw University Observatory, Aleje Ujazdowskie 4, 00-478 Warszawa, Poland}

\authorrunning{Naz\'e et al.}
\titlerunning{Periods for extraglactic Of?p stars}

\abstract{Galactic stars belonging to the Of?p category are all strongly magnetic objects exhibiting rotationally modulated spectral and photometric changes on timescales of weeks to years. Five candidate Of?p stars in the Magellanic Clouds have been discovered, notably in the context of ongoing surveys of their massive star populations. Here we describe an investigation of their photometric behaviour, revealing significant variability in all studied objects on timescales of one week to more than four years, including clearly periodic variations for three of them. Their spectral characteristics along with these photometric changes provide further support for the hypothesis that these are strongly magnetized O stars, analogous to the Of?p stars in the Galaxy. }


   \maketitle

\section{Introduction}
The spectral type Of?p, defined by \citet{wal72}, may at first sight seem awkward with its question mark, but it is not a typo at all. The ``f?p'' was introduced to identify stars showing strong emission lines of \niii\,$\lambda$\,4634-41 and \heii\,$\lambda$\,4686 (hence the ``f''), with a spectrum similar to those of supergiants, but showing other different characteristics including the unusual presence of \ciii\,$\lambda$\,4650 lines in strong emission, leading to their peculiarity flag (``p''); a question mark was then added to express doubt that these stars are normal Of supergiants. Only three Galactic objects were classified in that category in the seventies: HD\,108, HD\,191612, and HD\,148937 \citep{wal73}. 

In the last decade, a growing interest in this class has appeared, as recurrent spectral variations were identified in their Balmer and \hei\ lines. Derived periods amount to $\sim$7\,d for HD\,148937 \citep{naz08,naz10}, $\sim$538\,d for HD\,191612 \citep{wal04,naz07,how07}, and $\sim$55\,yrs for HD\,108 \citep{naz01,naz06}. A strong X-ray emission, brighter than that of normal O stars, was also found for these stars \citep{naz04,naz07,naz08}, with a modulation of the high-energy emission with the same period as the optical changes \citep{naz10}. Finally, optical photometric changes \citep{naz04th,bar07} as well as variations of line profiles in the UV range \citep{mar12,mar13} were also reported, and they display the same properties (same period, simultaneous extrema) as  the changes in the optical spectra.

\begin{table*}
 \centering
 \scriptsize
 \caption{List of our targets, with their position and their star and field IDs in ASAS and OGLE databases. Shifts applied to OGLE-II and OGLE-IV datasets to bring them in line with OGLE-III data are also provided (in bold).}
  \label{journal}
  \begin{tabular}{l c c c c c c}
  \hline
Star & RA & DEC & ASAS & OGLE-II & OGLE-III & OGLE-IV \\
  \hline
SMC159-2   & 00h49m58.72s & $-$73$^{\circ}$19'28.4'' &   & SMC\_SC5\#95303  & SMC100.8\#52967 & SMC720.27\#273 \\
& & & & {\bf ($\Delta=-$0.042\,mag)} & & {\bf ($\Delta=-$0.057\,mag)} \\
2dFS\,936  & 00h53m29.95s & $-$72$^{\circ}$41'44.5'' &  005329-7241.6, 005330-7241.6,  & SMC\_SC6\#237339 & SMC101.2\#21946 & SMC719.11\#68 \\
& & & 005331-7241.8, 005330-7241.7, & {\bf ($\Delta=-$0.013\,mag)} & & {\bf ($\Delta=-$0.032\,mag)}\\
  & &  &   and 005329-7241.4 & & \\
AV\,220    & 00h59m09.97s & $-$72$^{\circ}$05'48.3'' &  005910-7205.8 &                                             & SMC108.3\#6501 & SMC725.16\#45 \\
& & & and 005909-7205.8 & & & {\bf ($\Delta=+$0.006\,mag)}\\
BI\,57     & 05h01m08.59s & $-$68$^{\circ}$11'45.1'' &  050107-6811.6, 050109-6811.8, & & LMC125.2\#9 & LMC510.32\#25733 \\
& & & 050111-6811.6, 050106-6811.6,  & & & {\bf ($\Delta=+$0.040\,mag)} \\
   &  &  &  and 050108-6811.8 &                                             &  & and LMC532.08\#12629 \\
& & & & & & {\bf ($\Delta=+$0.026\,mag)}\\
LMC164-2   & 05h13m49.88s & $-$69$^{\circ}$23'21.7'' &  051351-6923.1, 051346-6923.1, & LMC\_SC9\#216949 & LMC111.2\#10462 & LMC503.15\#28081 \\
& & &051349-6923.4, 051351-6923.1, & {\bf ($\Delta=-$0.003\,mag)} & & {\bf ($\Delta=+$0.013\,mag)}\\
& & &051350-6923.4, 051350-6923.1\\
\hline
\end{tabular}
\\
Notes: 2dFS\,936 is also known under the name [MD2001] Anon 1; 
ASAS dataset 051346-6923.1 was discarded as it contains only two flag 'A' points and in one of them the star appeared much brighter than in all other data.\\
\end{table*}

After a strong magnetic field was detected in HD\,191612 \citep{don06}, the magnetic oblique rotator scenario was proposed to explain these peculiar properties. The monitoring of \citet{wad11} confirmed the proposed scenario for that star. In this model, the stellar winds of massive stars are confined in the equatorial regions by a strong dipolar magnetic field and, as the star rotates, these regions come in and out of direct view, generating the observed recurrent variations found at all wavelengths. In this context, the recurrence timescale is the rotational period, and long rotation periods may then be explained by magnetic braking over the lifetime of the star \citep{udd09}. This implied that all Of?p stars are strongly magnetic, and HD\,108 and HD\,148937 were indeed soon found to be so \citep{hub08,hub11,hub13,mar10,wad12}. 

At the same time, new Of?p stars were identified in the Galaxy, with very similar characteristics \citep{wal10}: NGC1624-2 \citep[period of 158\,d,][]{wad12dash} and CPD$-28^{\circ}$2561 (\citealt{hub11,hub13}; period of 73\,d, \citealt{wad15}). It must be stressed that, while about ten O stars are now known to be magnetic, the only actual class of magnetic massive stars is the Of?p category. Spotting the typical spectral properties of Of?p stars therefore constitutes a powerful, albeit indirect, way to identify a magnetic O-type star. However, some stars with spectral types O3.5--5.5 actually display strong \ciii\ lines naturally (i.e. in the absence of a magnetic field), which led to the definition of a separate spectral type dubbed ``Ofc'' \citep{wal10,marhil12}. The presence of strong \ciii\ lines is therefore not sufficient for classifying a star as Of?p. Moreover, it is now known that the \ciii\ emission lines in some Of?p spectra disappear entirely at certain phases, while in others they remain strong or weak at all phases. Hence, the identification of an Of?p-type star is now rather performed through a large body of evidence: peculiar spectral characteristics (e.g. in the optical: emission lines narrower than absorption lines, emissions or P Cygni features within Balmer and \hei\ lines,etc.) that are often periodically variable, small recurrent variations in brightness, abnormally bright X-ray emissions, and strong dipolar magnetic fields \citep[see e.g.][]{naz08b}. 

Like magnetic fields, metallicity is a key ingredient for many aspects in massive stars' lives (formation, evolution, winds). It is therefore of utmost interest to study whether magnetism also occurs, and with what characteristics, at other metallicities. The closest laboratories for such tests are the Magellanic Clouds (MCs), and the best way to find likely magnetic stars is to locate Of?p stars. Indeed, other kinds of magnetic stars, e.g. active M dwarfs or Ap stars \citep{mai01,pau11}, might be easier to analyse, but the much higher intrinsic brightness of O stars makes them better targets in the MCs. In fact, peculiarities typical of Galactic Of?p stars have already been detected for five stars in the MCs: AV\,220 \citep{wal00,mas01}, 2dFS\,936 \citep{mas01,eva04}, BI\,57 (I. Howarth, private communication; Walborn et al. in preparation), SMC159-2 and LMC164-2 \citep{mas14}. A sixth one, LMC N82 (or Brey 3a) was proposed by \citet{hey92}, but it was classified as Ofp by \citet{wal03}. Nothing more is currently known about these five stars, as their great distance makes the investigations difficult. This paper provides the first step forward, with the analysis of photometric data of these stars, with the aim of identifying a key characteristic of magnetic Of?p stars: periodic variability consistent with a long (weeks-years) rotational period. Section 2  presents the observations used to do so, while Section 3 presents our results, and Section 4 concludes this paper.

\section{Data}
In Galactic Of?p stars, periods range from a week to decades. Trying to pinpoint the recurrence timescale from random spectral observations can thus be very time-consuming. It is even more the case when the faintness of the object imposes long exposures to get spectra of good quality, as for MC objects. However, Galactic Of?p stars do not only display spectral variations, they also show photometric changes, of the order of 0.05\,mag for HD\,108 and HD\,191612 \citep{koe02,naz04th,bar07}. Brightness variations are much easier to detect for MC objects since there have been several photometric surveys of these regions over the years. 

We thus searched the archives of the OGLE-II \citep{uda97}\footnote{http://ogle.astrouw.edu.pl/} and ASAS \citep{poj97}\footnote{http://www.astrouw.edu.pl/asas/} projects. We found data (in the $I$-band for OGLE-II, $V$-band for ASAS) for the five candidate Of?p stars in the MCs. From these datasets, only the best quality data were kept (box ``good quality'' selected for OGLE-II when downloading data - we note, however, that data with flag down to F remain; only data with flag A were kept for ASAS). For OGLE-II, two photometric datasets are available: one called DIA which refers to Difference Image Analysis \citep{ala98,woz00}, and one called PSF which refers to PSF fitting \citep{szy05}. The analyses were performed on both sets of magnitudes: they gave similar results. As DIA data are less noisy (and known to be more reliable, see \citealt{szy05}), the numbers presented below refer to them. 

Further OGLE data, from OGLE-III and OGLE-IV campaigns, were also available. The third phase of the OGLE project (OGLE-III, \citealt{uda03}) was realized during the years 2001-2009 using an 8-chip CCD mosaic. The photometry was obtained using the DIA method. OGLE-IV is the currently ongoing, fourth phase of the project. It started in 2010, using a huge, 32-chip CCD mosaic and the DIA reduction scheme. Analysing the OGLE datasets we noted small residual magnitude shifts between the three campaigns. To correct for them, we computed, for each dataset, the mean of all data, the median of the 10\% largest data (i.e. 90th percentile), and the median of the 10\% smallest data (i.e. 10th percentile). OGLE-III data are considered as the best calibrated amongst the OGLE collaboration, and were then used as reference: the differences between the derived mean and median values of OGLE-II/IV and their equivalent in OGLE-III data yield shifts that were applied to the  OGLE-II/IV before analysis. We note that using means or any of the medians yields the same shifts. For reference, they are shown in Table \ref{journal} along with the IDs of the fields and stars in the different programmes. 

Although it is currently off-line, we managed to get access to the EROS-2 database (see \citealt{tis07} for details). Data are available for three of our stars (AV\,220, 2dFS\,936, and LMC164-2). They cover a shorter time interval than OGLE data, from HJD$\sim$2\,450\,300 to 2\,452\,700, but can be used as a confirmation tool. Other photometric surveys have no available data for these objects (e.g. Catalina sky survey) or have no public data. 

\begin{table}
 \centering
 \scriptsize
  \caption{Properties of the detected periods, from the Fourier method.}
  \label{list}
  \begin{tabular}{lccc}
  \hline
ID & $T_0-$2\,450\,000 & period (d) & amplitude (mag)\\
  \hline
SMC159-2   & 3598.706$\pm$0.049& 14.914$\pm$0.004     & 0.00719$\pm$0.00015 \\
2dFS\,936$^{\dagger}$  & 3993.3$\pm$1.7    & 1370$\pm$30          & 0.01504$\pm$0.00011 \\
LMC164-2   & 4027.338$\pm$0.015& 7.9606$\pm$0.0010    & 0.00926$\pm$0.00011 \\
BI\,57$^*$ & 5302.2$\pm$1.2    & 400.0$\pm$3.5        & 0.00550$\pm$0.00010 \\
\hline
\end{tabular}
\\
$T_0$ (hence $\phi=0$) corresponds to minimum brightness; $^*$ not strictly periodic (see text); $^{\dagger}$there are additional periodicities in 2dFS\,936 (see text).\\
\end{table}

On each of these datasets, we first applied a $\chi^2$ test for constancy and then several period search algorithms. They fall in four categories: (1) the Fourier algorithm adapted to sparse/uneven datasets \citep[a method rediscovered recently by \citealt{zec09} - these papers also note that the method of \citet{sca82}, while popular, is not fully correct, statistically]{hmm,gos01}, (2) two different string length methods \citep{lafkin,renson}, (3) three binned analyses of variances (\citealt{whi44}, \citealt{jur71} which is identical, with no bin overlap, to the ``pdm'' method of \citealt{ste78}, and \citealt{cuy87} - which is identical to the ``AOV'' method of \citet{sch89}), and (4) conditional entropy \citep[see also \citealt{gra13}]{cin99,cin99b}. Each of these methods has its advantages and its drawbacks - the most reliable being the Fourier method while the fastest ones usually are analyses of variances - but a period clearly identified by all of them is certainly real. All uncertainties reported in this paper correspond to 1$\sigma$ errors.

\begin{figure}
\centering
\includegraphics[width=8cm]{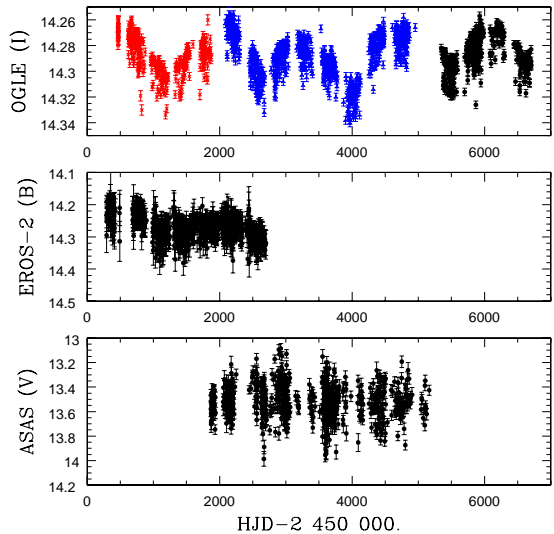}
\caption{Photometric variations of 2dFS\,936, observed in the I-band (OGLE, top - OGLE-II data are shown by red crosses, OGLE-III by blue triangles, and OGLE-IV by black dots), non-standard B-band (EROS-2, middle), and V-band (ASAS, bottom). It must be noted that the top panel spans only 0.10\,mag while the middle and bottom panels span much larger ranges (0.4\,mag and 1.2\,mag, respectively).}
\label{2dfs}
\end{figure}

\begin{figure*}
\includegraphics[width=6cm]{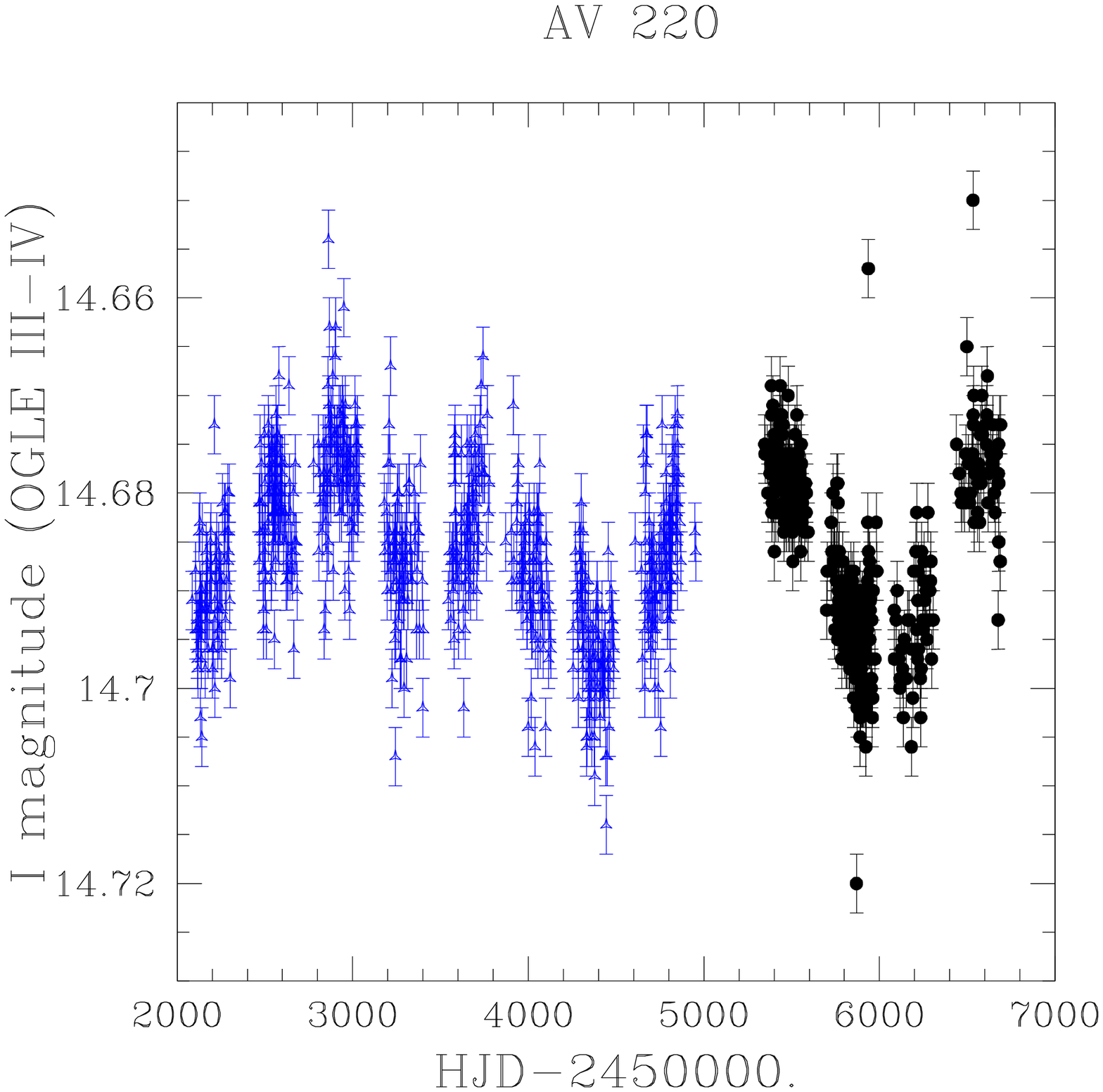}
\includegraphics[width=6cm]{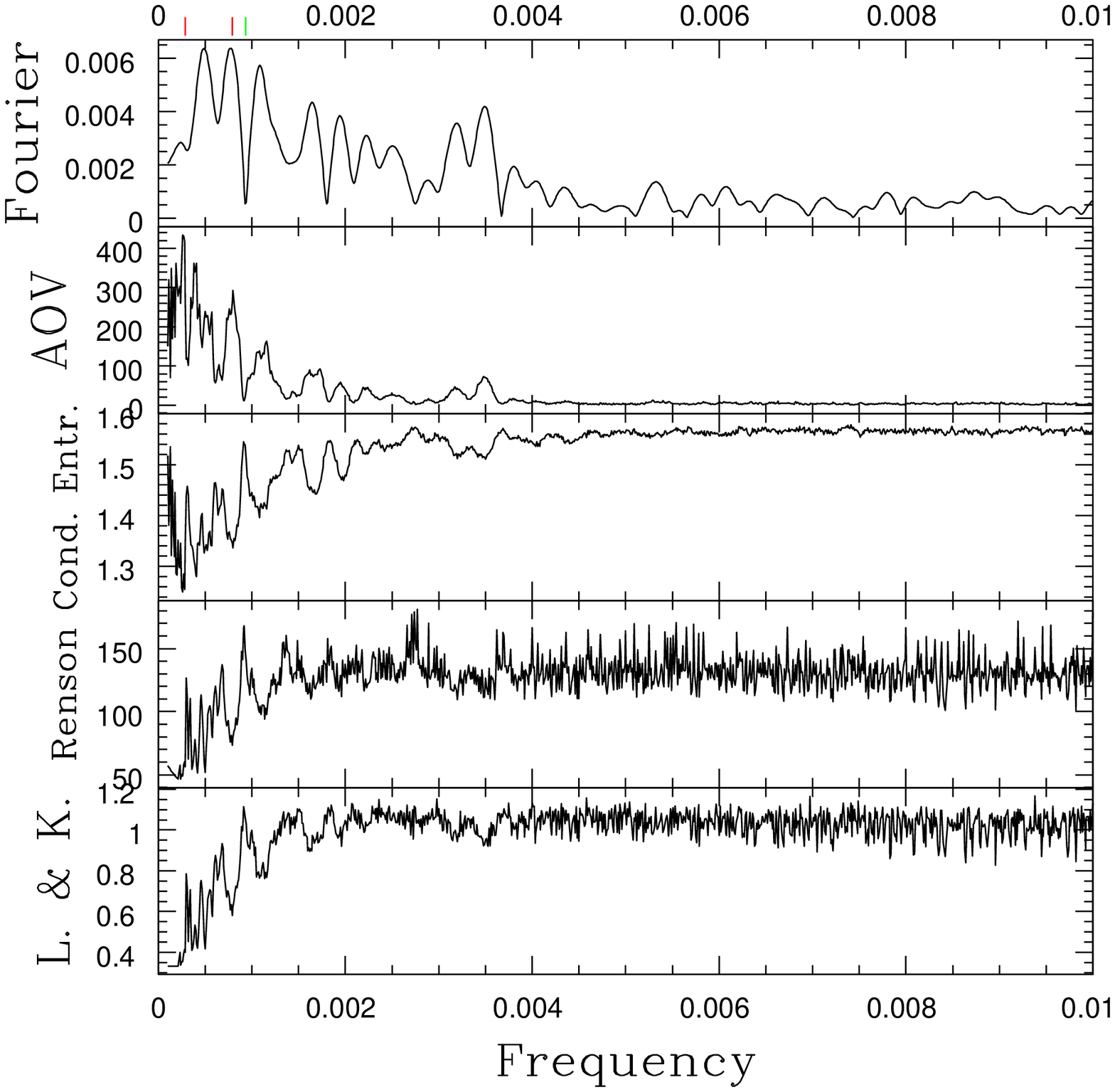}
\includegraphics[width=6cm]{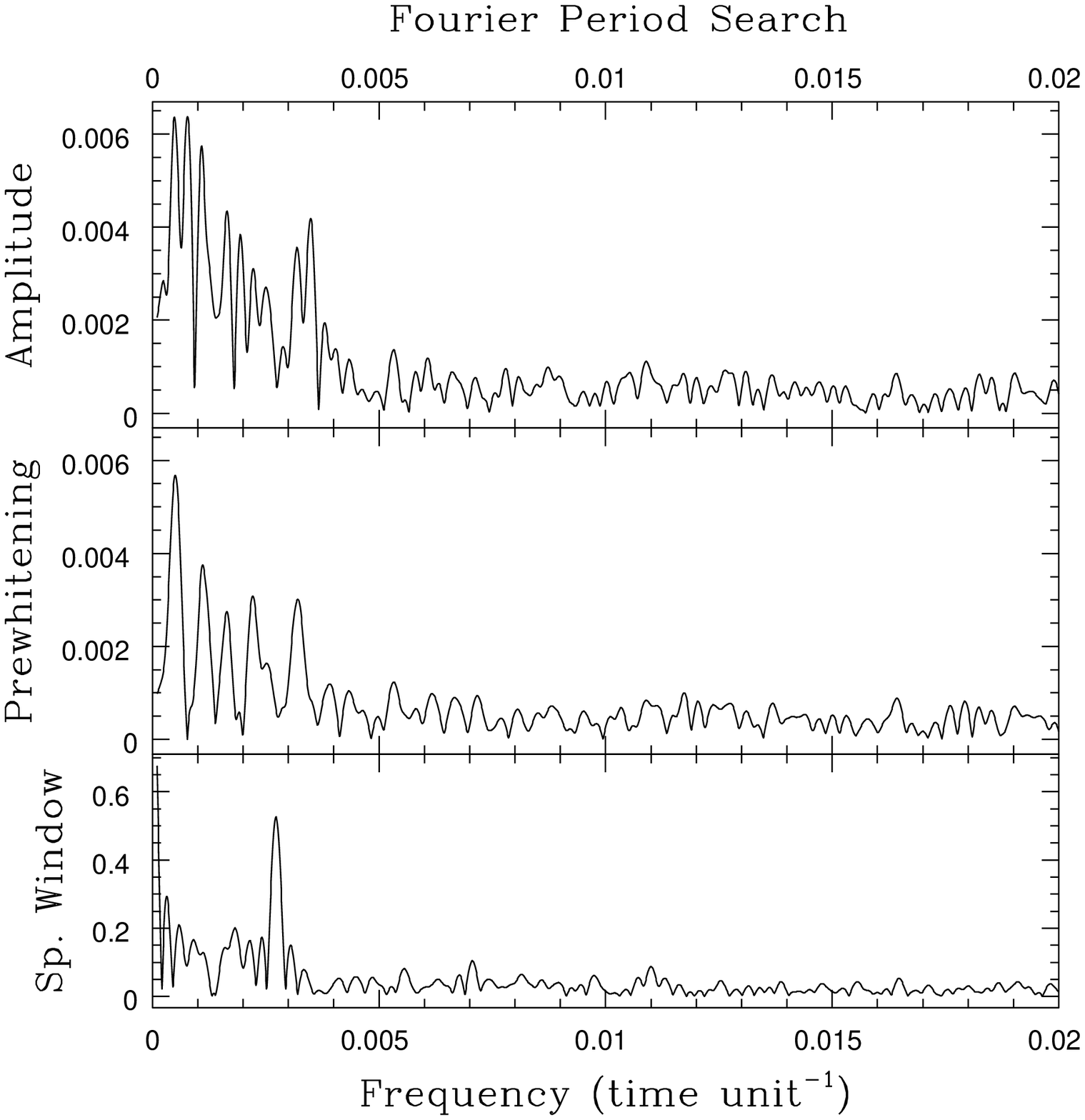}
\caption{{\it Left:} OGLE-III and IV lightcurve for AV\,220 (colour scheme as in Fig. \ref{2dfs}). {\it Middle:} Comparison of the different period search methods, with tickmarks at the top indicating periods of 3846\,d, 1299\,d, and 1100\,d. The presence of a signal is marked by a maximum for Fourier and AOV methods, but by a minimum for conditional entropy, Renson and Lafler \& Kinman methods. {\it Right:} Fourier periodogram for the raw (top) and prewhitened (middle, for the best-fit Fourier period $P$=1299\,d) data, along with the spectral window (bottom). }
\label{av}
\end{figure*}

\begin{figure*}
\includegraphics[width=6cm]{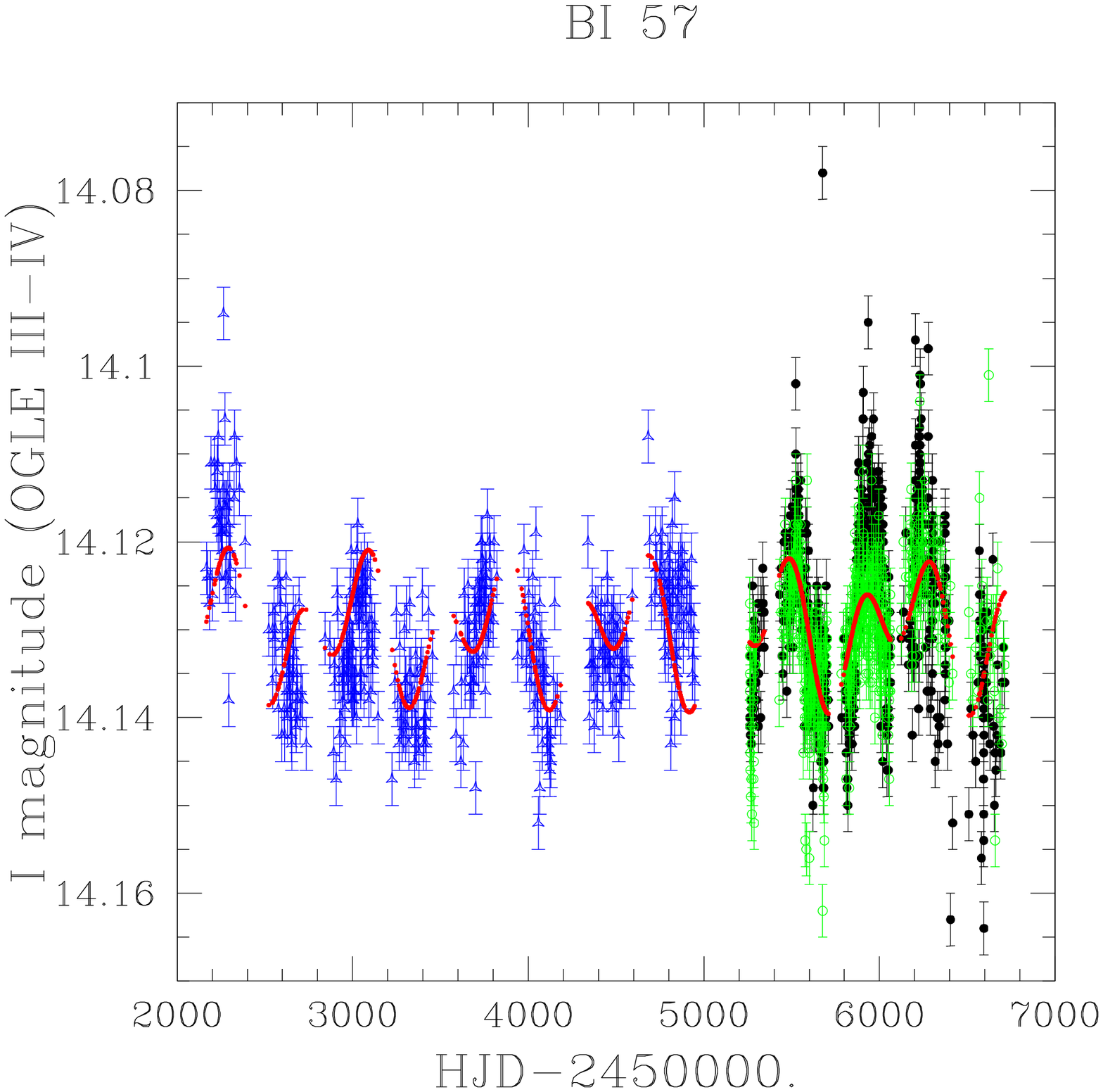}
\includegraphics[width=6cm]{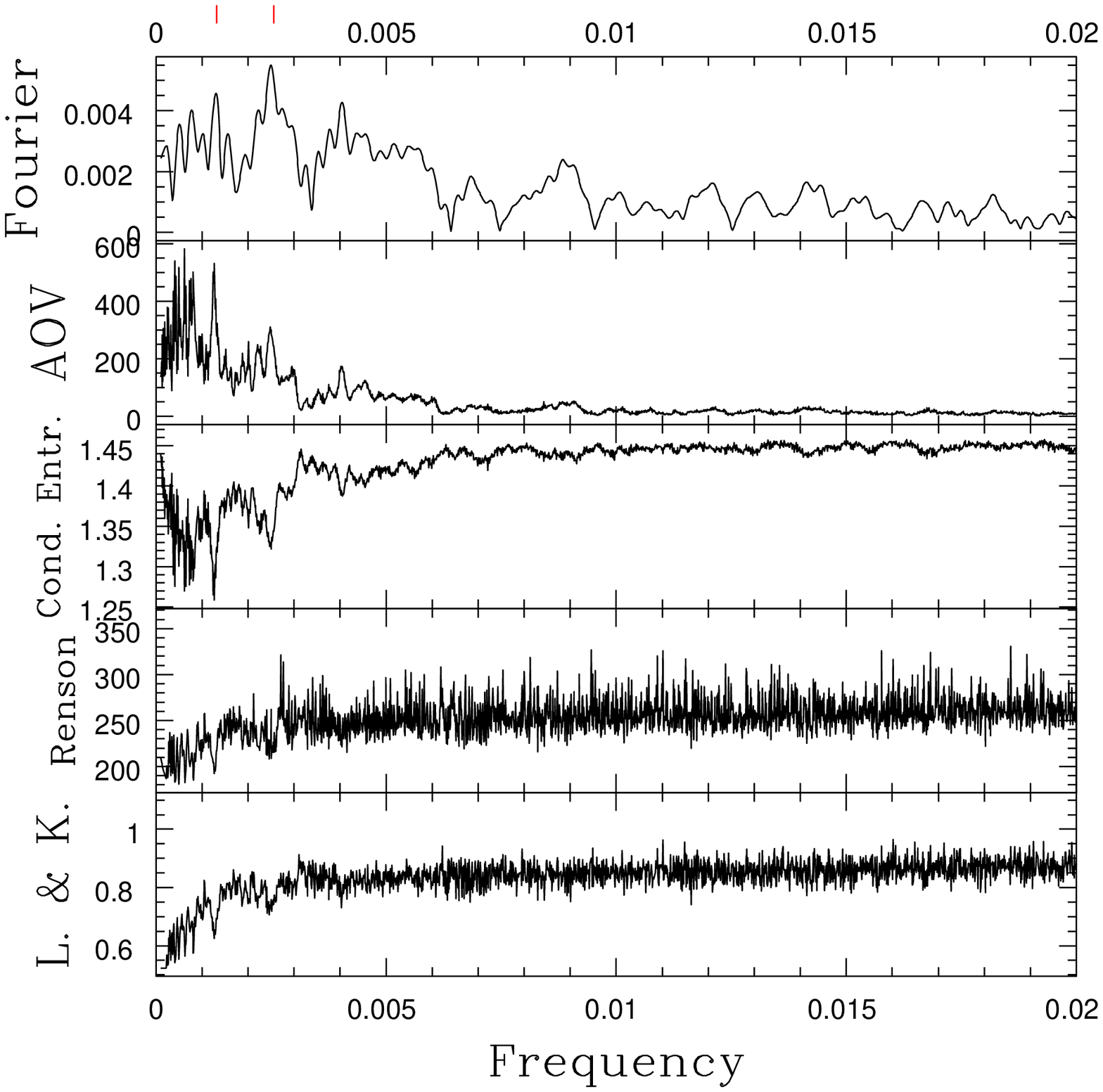}
\includegraphics[width=6cm]{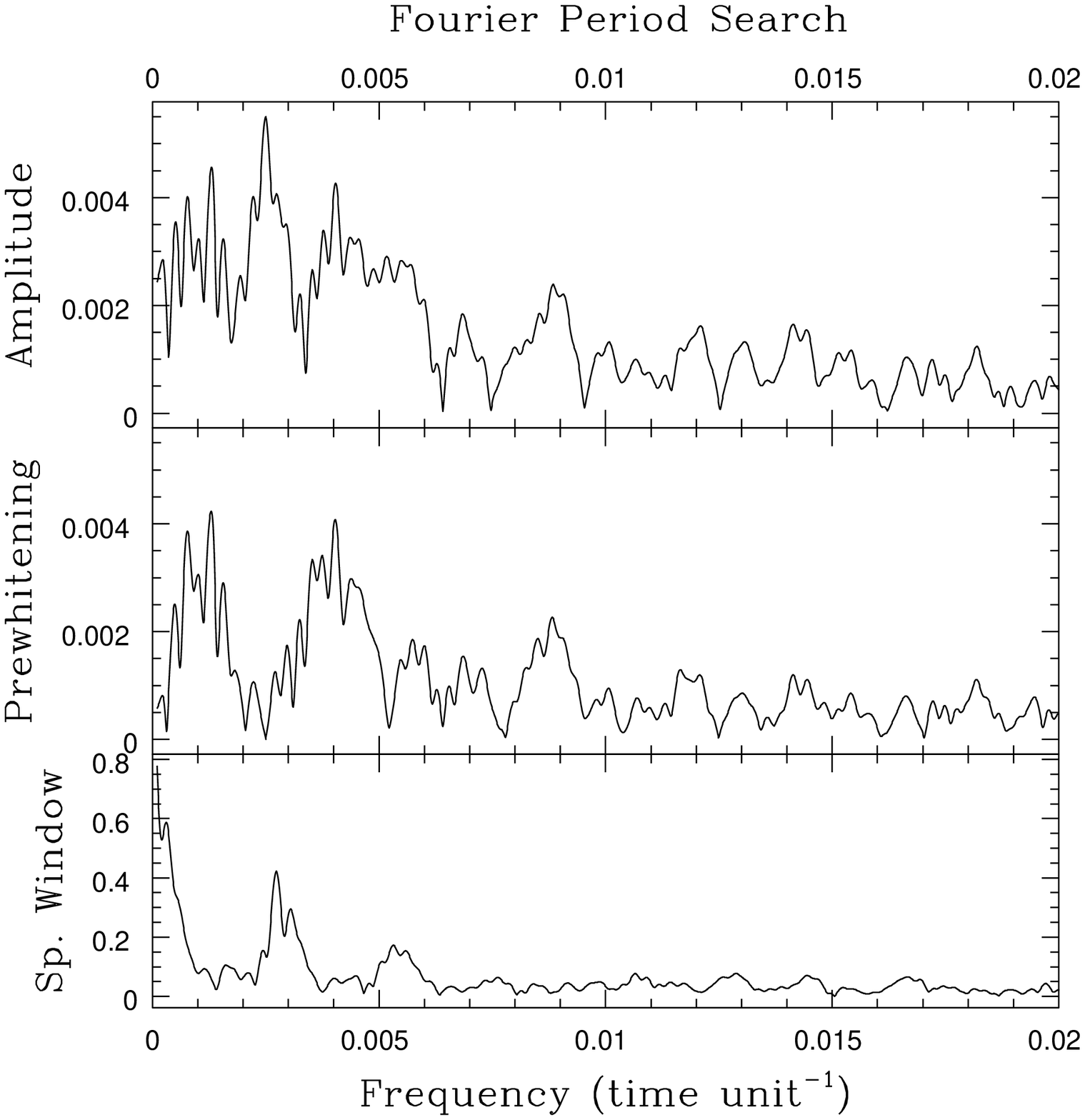}
\caption{Same as Fig. \ref{av} for BI\,57 (a best-fitting double sinusoid, with the two periods of 787\,d and 400\,d is added in red in the left panel; these periods are indicated by tickmarks in the middle panel; prewhitening in the right panel was performed for $P$=400\,d). The potentially problematic OGLE-IV dataset (LMC532.08) is shown in green.}
\label{bi}
\end{figure*}

In view of the results (see next section), additional spectra are needed. Indeed, usually, only one spectrum for each object exists in the literature, but more than one spectrum is needed to examine variations. Spectral monitorings of AV\,220, BI\,57, and 2dFS\,936 are examined in a companion paper (Walborn et al., in preparation). For the remaining two objects, the published spectra of LMC164-2 and SMC159-2 were obtained in the framework of a WR survey in the MCs \citep{mas14}. This campaign is not completed yet, which enabled us to get one additional spectrum of SMC159-2 in September 2014. As it is the sole new one, it will be presented here rather than by Walborn et al. (in preparation) who present larger spectral sets. As with the two previous spectra \citep{mas14}, this new spectrum was obtained at Las Campanas Observatory with the 6.5-m Magellan II (Clay) telescope using the MagE spectrograph \citep{mar08} and a 1'' slit which provided a resolution ranging from 0.7 to 1.7\,\AA\ going from the blue to the red end of the optical spectrum. The exposure time was 3$\times$600s. The data were processed with a combination of the IRAF {\it mtools} package, originally developed by Jack Baldwin for the reduction of MIKE data, and standard IRAF \'echelle tasks as described by \citet{mas12}. ThAr lamps were used to derive wavelength solutions, and flux standards were observed on each night in order to flux calibrate the data. The two spectra of SMC159-2 were normalized considering clean continuum windows.

\section{Results}
Before looking at each object in turn, a note about the data quality should be made. We expect the photometric variations to be very small (in the Galaxy, variations no larger than 0.05\,mag are seen, see previous section). OGLE data typically have errors of 0.003-0.004\,mag while the errors on ASAS and EROS-2 data are ten times larger (0.04--0.05\,mag for ASAS and 0.015--0.15\,mag with a median of 0.03\,mag for EROS-2). In addition, the extraction apertures are different (from about 1.5'' in OGLE to a radius of 15'' in ASAS, with EROS-2 in-between), leading to different degrees of contamination. It is therefore unsurprising that the different datasets do not produce exactly the same results. A good example is provided with 2dFS\,936: the coherent variations detected by OGLE have an amplitude of 0.015\,mag and this is confirmed in EROS-2 data (see below), but this signal is buried in ASAS noise as can be seen from a direct comparison of data taken by the two surveys at similar epochs (Fig. \ref{2dfs}). Therefore, we definitely trust more OGLE data than ASAS data.

\subsection{AV\,220}
For AV\,220, the photometric data show a dispersion that is significantly larger than the error bars (e.g. Fig. \ref{av}, left panel), and are thus found significantly variable in a $\chi^2$ test. However, period search algorithms do not find a clear and unequivocal signal (Fig. \ref{av}, middle panel). With the ASAS dataset, most methods display some low-amplitude peaks around 10--15\,d, but periodograms look overall like white noise. For OGLE data, power is found only at the lowest frequencies, but different methods disagree on the value of the best period (1299\,d, 3846\,d?), and prewhitening with these frequencies does not yield a flat periodogram (Fig. \ref{av}, right panel). This may come from a changing variation timescale, with a variation apparently more rapid in the OGLE-IV dataset. Indeed, when looking at photometry over time (Fig. \ref{av}, left panel), brightness maxima are seen around $HJD=$2\,452\,900, 2\,455\,200, and 2\,456\,600 while minima are seen around $HJD=$2\,452\,000, 2\,454\,400, and 2\,456\,000 (we note, however, that 2\,455\,200 and 2\,456\,600 are at the beginning and end, respectively, of the dataset). To assess this difference, we perform period searches on the OGLE-III and OGLE-IV datasets separately: periodograms based on OGLE-III data favour ``periods'' of 2500\,d or 3700\,d, while those calculated with OGLE-IV data yield shorter values of 1200--1400\,d. The change with time is thus not an effect of the annual gaps, but a true variation. The EROS-2 data produce similar results and could not provide a stronger constraint on the timescales. A secure period has thus not yet been identified for AV\,220, and further monitoring is needed to better understand the photometric variations of this star.

\subsection{BI\,57}
\begin{figure}
\includegraphics[width=8.5cm]{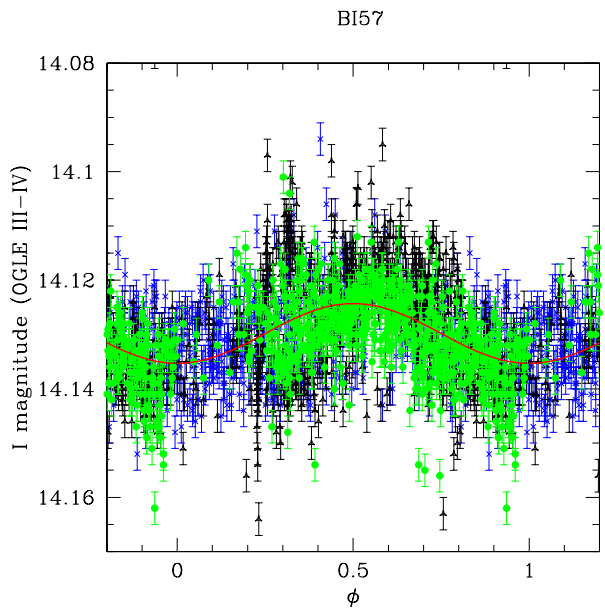}
\caption{Photometric variations of BI\,57, phased using the best-fit Fourier ephemeris (period of 400\,d, see Table \ref{list}). The thick red line corresponds to the best-fit Fourier sinusoid for that period, while the symbols for data are as in Fig. \ref{bi}. }
\label{biph}
\end{figure}

Regarding BI\,57, it must be noted that, in the LMC532.08 field, the star is located close to the outer edge of the mosaic, where the template image is of worse quality and the photometry calibration less reliable owing to vignetting. However, discarding this dataset from the analysis does not change significantly the following results. BI\,57 displays obvious and significant variability of its photometry (Fig. \ref{bi}, left panel). In period searches, ASAS data yield a marginal detection at a Fourier period of $P=96.7\pm0.3$\,d, while OGLE data clearly favour a period of 400$\pm$3\,d (Fourier) or its double (787$\pm$14\,d, entropy \& variance analyses). While the photometry appears rather well phased with the latter periods (e.g. Fig. \ref{biph}), it must be noted that the dispersion in the phased diagram appears higher than for the three following Of?p candidates, pointing to incompatibilities of the OGLE dataset with a strict periodicity. As further evidence, a model considering the sum of two sinusoids with periods 400\,d and 787\,d does reproduce the average magnitudes of each observing run, but sometimes fails to reproduce the detailed trends (increase/decrease) of the observed variations (see such a model in red in the left panel of Fig. \ref{bi}). Furthermore, prewhitening by these periods does not yield a flat periodogram as it should in the case of a strict periodicity (e.g. Fig. \ref{bi}, right panel). Analysing the three OGLE datasets separately confirms the presence of differences: for example, the 787\,d peak is higher in the Fourier periodogram of OGLE-III data while the periodograms based on the two OGLE-IV datasets favour the 400\,d peak. Therefore, while it is clear that some changes occur with periods of 400\,d and 787\,d, such periodicities cannot explain the full range of variations in BI\,57.

\subsection{2dFS\,936}
\begin{figure*}
\includegraphics[width=6cm]{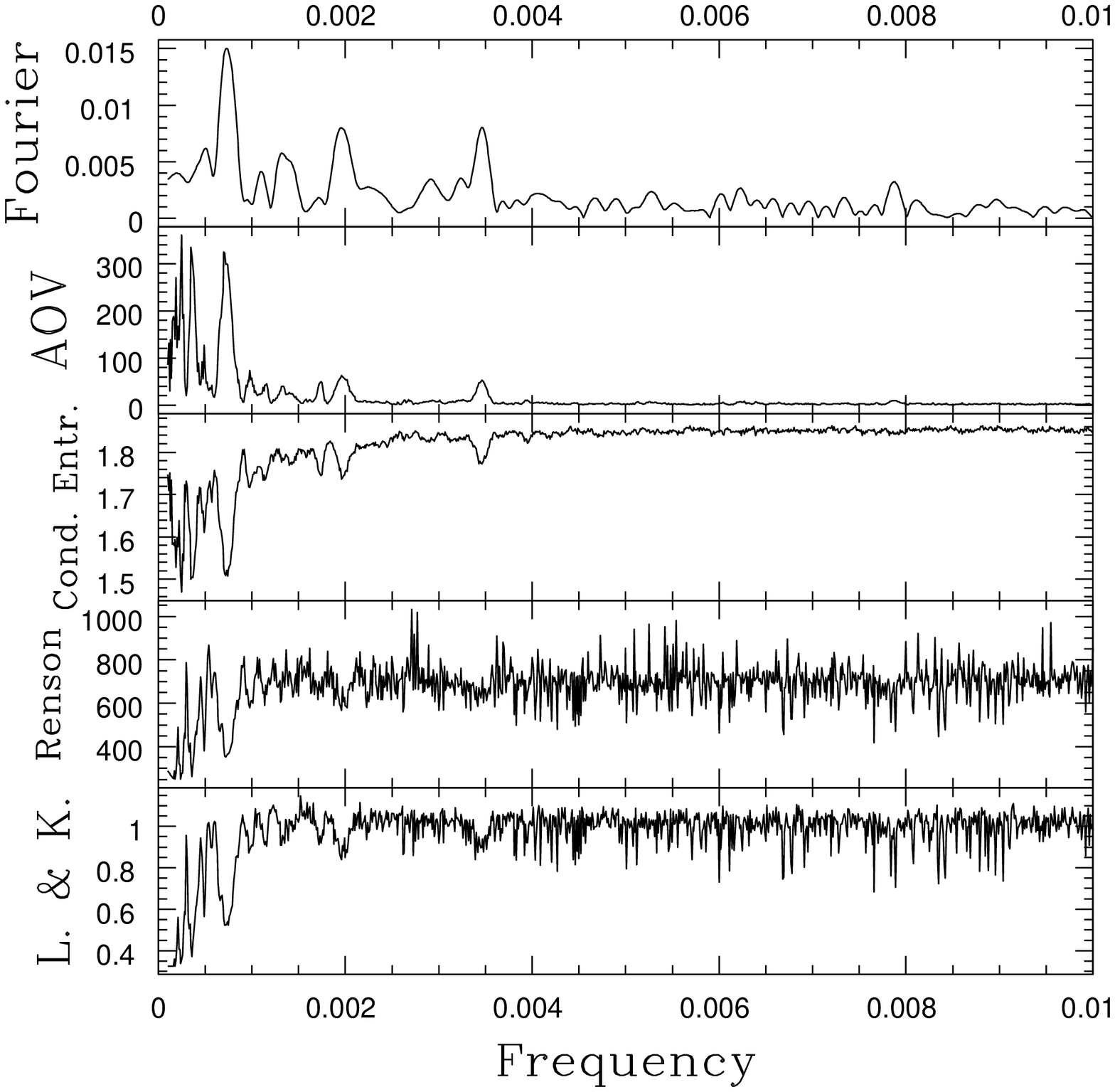}
\includegraphics[width=6cm]{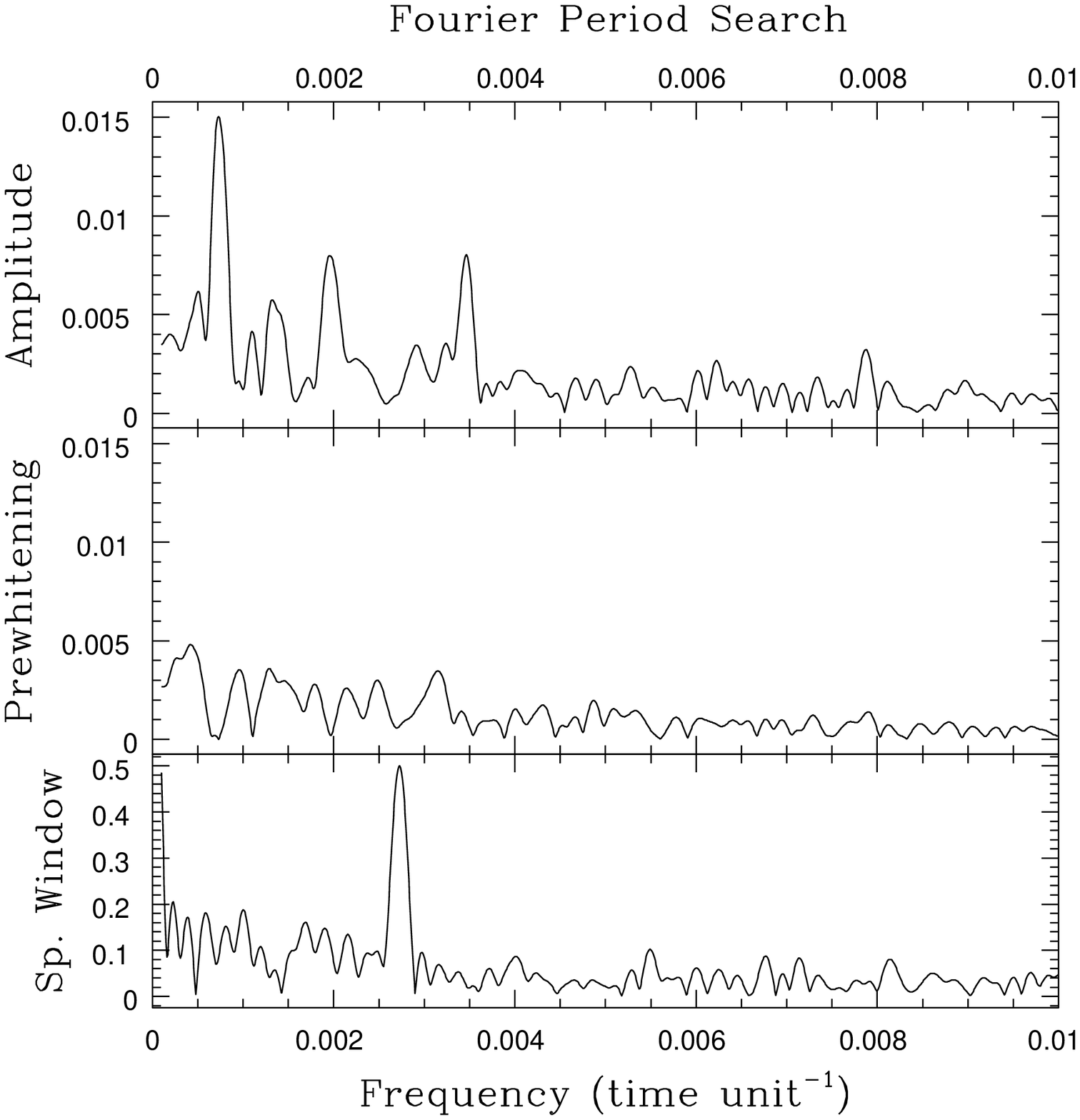}
\includegraphics[width=6cm]{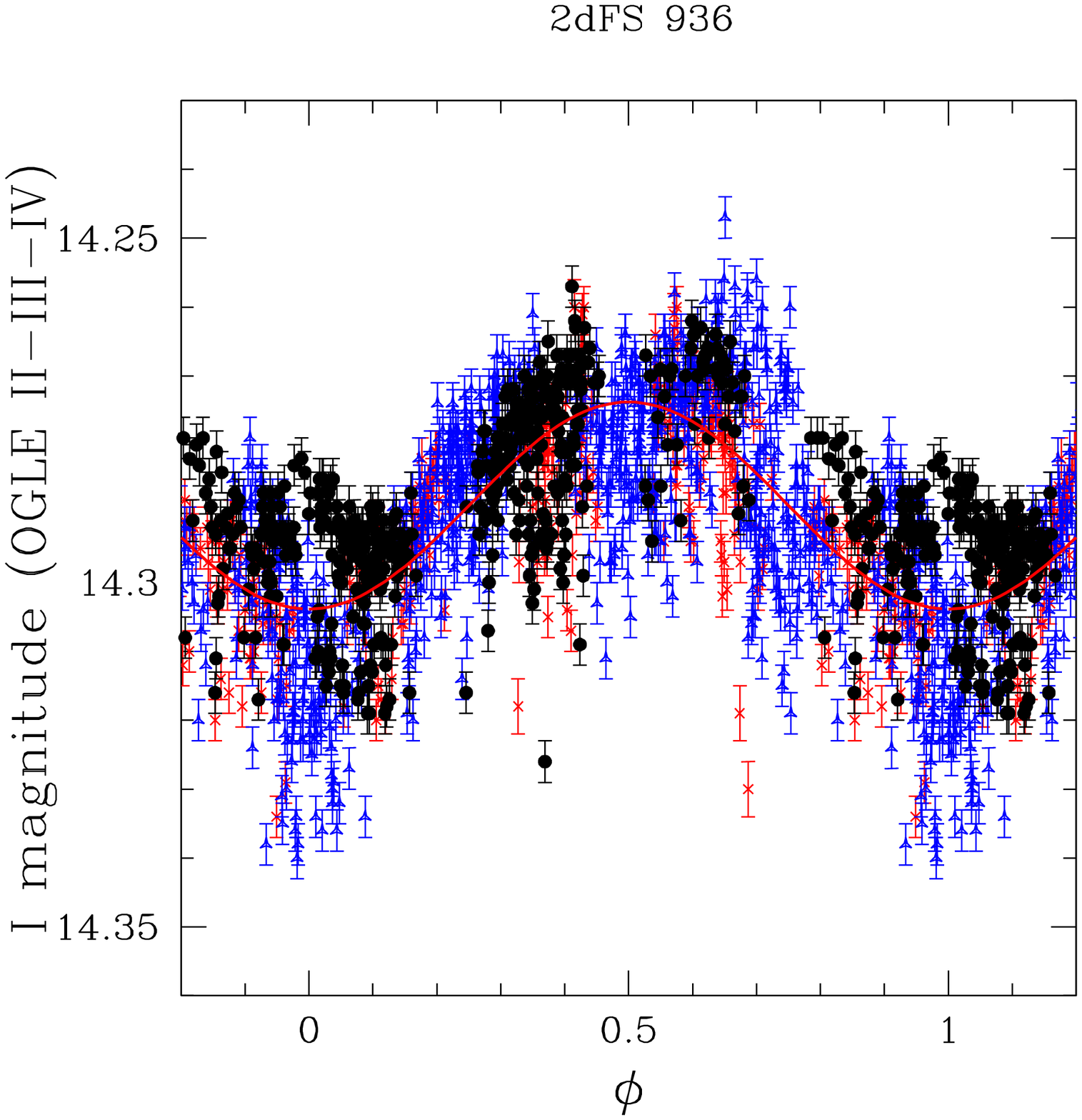}
\caption{Period search results for 2dFS\,936. {\it Left:} Comparison of the different period search methods. The presence of a signal is marked by a maximum for Fourier and AOV methods, but a minimum for conditional entropy, Renson and Lafler \& Kinman methods. {\it Middle:} Fourier periodogram for the raw (top) and prewhitened (middle, for the best-fit Fourier period) data of 2dFS\,936, along with the spectral window (bottom). {\it Right:} Photometric variations of 2dFS\,936, phased using the best-fit ephemeris (see Table \ref{list}). The thick red line corresponds to the best-fit Fourier sinusoid, while the symbols for data are as in Fig. \ref{2dfs}.}
\label{2dfsbis}
\end{figure*}

The case of 2dFS\,936 is very different. The data appear significantly variable (the star was in fact reported as an ``irregular variable'' by \citealt{kou14}); a clear long-term modulation is readily detected by eye in OGLE data (Fig. \ref{2dfs}). It is indeed detected independently by the different period search methods (Fig. \ref{2dfsbis}, left panel), with a best-fit period of 1370$\pm$30\,d (Table \ref{list}). Prewhitening by this period yields a flat periodogram while phased photometry results in very coherent variations (Fig. \ref{2dfsbis}, middle and right panels), securing that detection. In addition, the spectroscopic changes of 2dFS\,936 seem to occur in phase with the same period (see Walborn et al., in preparation): this 1370\,d period clearly appears as the rotation period of a typical Of?p star. 

\begin{figure}
\includegraphics[width=8.5cm]{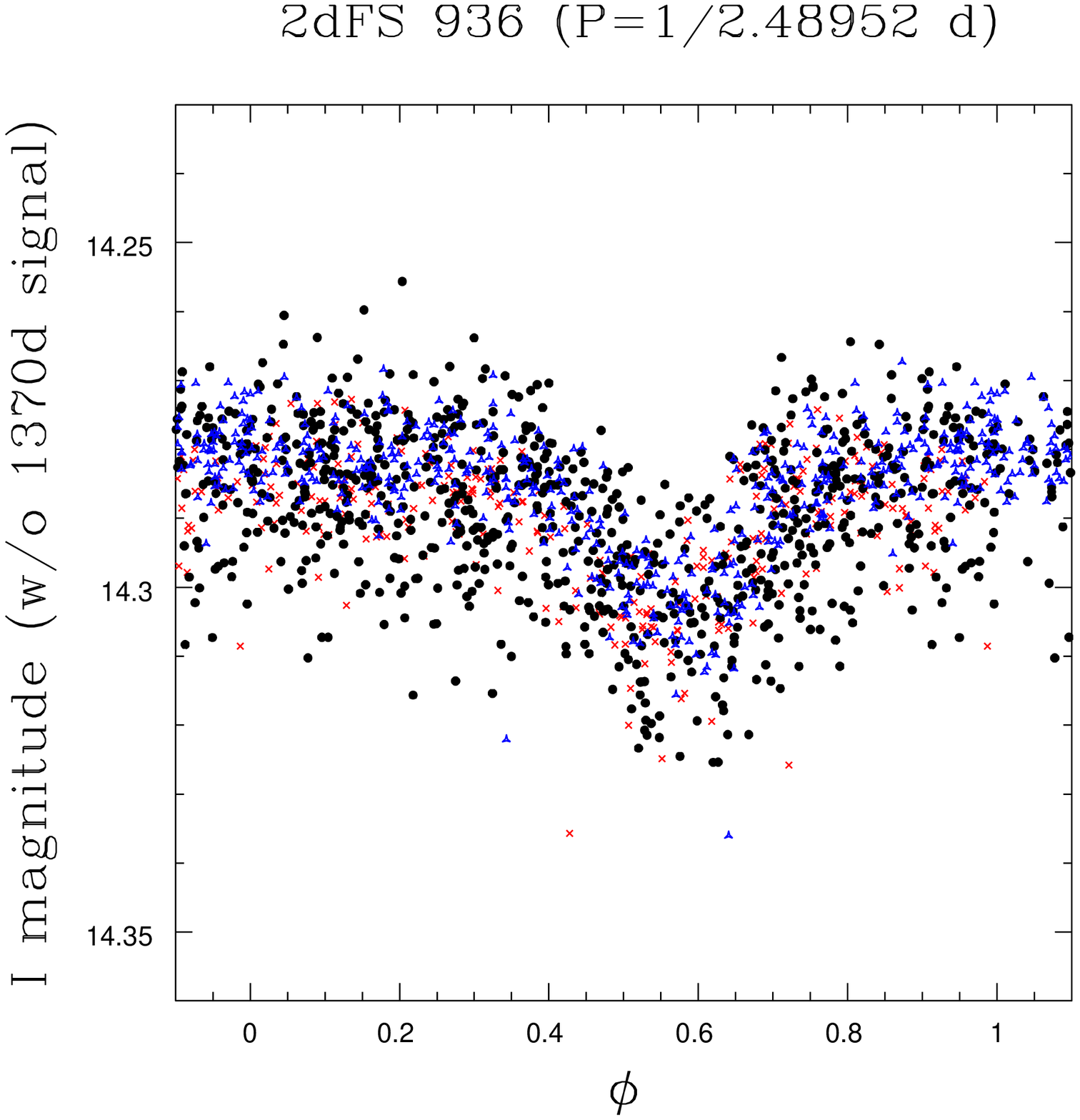}
\caption{Photometric changes of 2dFS\,936, once the 1370\,d variation is removed. This plot uses the $T_0$ of Table \ref{list} and a period of 0.402\,d (i.e. a frequency of 2.48952\,d$^{-1}$). The symbols for data are as in Fig. \ref{2dfs}. }
\label{2dfsph}
\end{figure}

It should be noted, however, that the maximum at $HJD\sim$2\,453\,250 and the following minimum at $HJD\sim$2\,454\,000 occur when the star was fainter than for other maxima and minima. Examining further the periodogram, we found that a second set of peaks exists, along with their daily aliases: in the interval 0.--4.5\,d$^{-1}$, their frequencies are 0.48525\,d$^{-1}$ \& 0.51246 \,d$^{-1}$, 1.48676\,d$^{-1}$, 2.48952\,d$^{-1}$, 3.48952\,d$^{-1}$ \& 3.49224\,d$^{-1}$, and 4.49228\,d$^{-1}$. With the current dataset, it is difficult to assess which frequency is the correct one: data points prewhitened for the 1370\,d variation yield nicely phased variations for all frequencies larger than 1\,d$^{-1}$ (e.g. Fig. \ref{2dfsph}). The origin of this additional variation may not be linked to the Of?p phenomenon, it may rather be related to pulsations or to eclipses in a binary system. In the latter case, the orbital period would then be the double of one of the periods detected by period searches and the lightcurve with two similar eclipses would indicate similar stars. However, such a period (0.4--4\,d) would be very short for such massive objects and a line-of-sight coincidence between 2dFS\,936 and an eclipsing binary could then be required. Identifying the correct frequency and understanding its nature implies reobserving 2dFS\,936, this time over several nights with a high sampling frequency to avoid any ambiguities. 

Formally, using a $\chi^2$ test, the EROS-2 data in their blue band (where most data points are available) do not deviate significantly from a constant, but period search methods, being very sensitive to periodic signals, readily confirm the signals found in OGLE data: main period at 1389$\pm$80\,d with an amplitude in B-band of 0.0206$\pm$0.0013\,mag, secondary period at 2.05\,d (or its aliases).

\subsection{LMC164-2}
\begin{figure*}
\includegraphics[width=6cm]{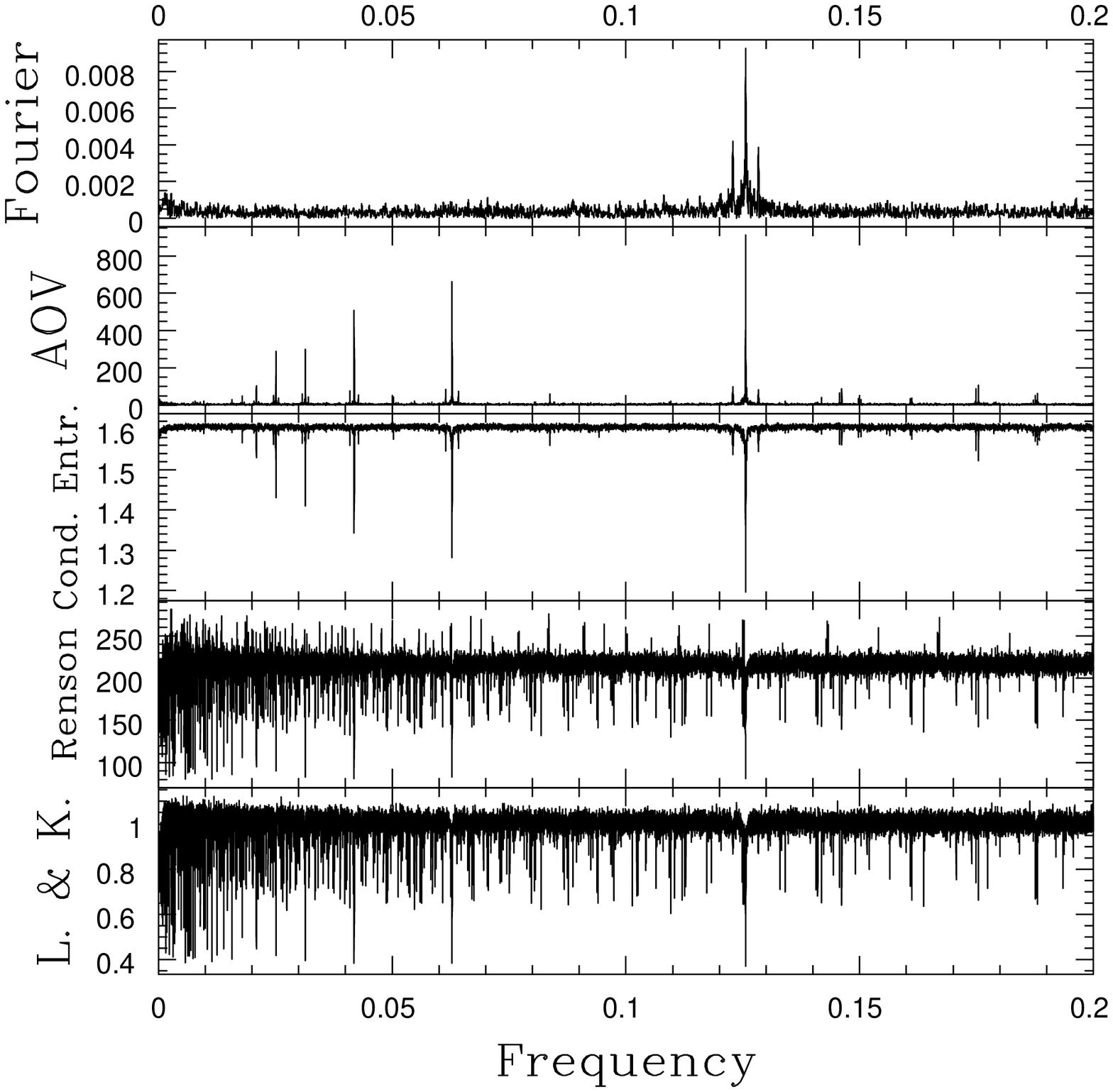}
\includegraphics[width=6cm]{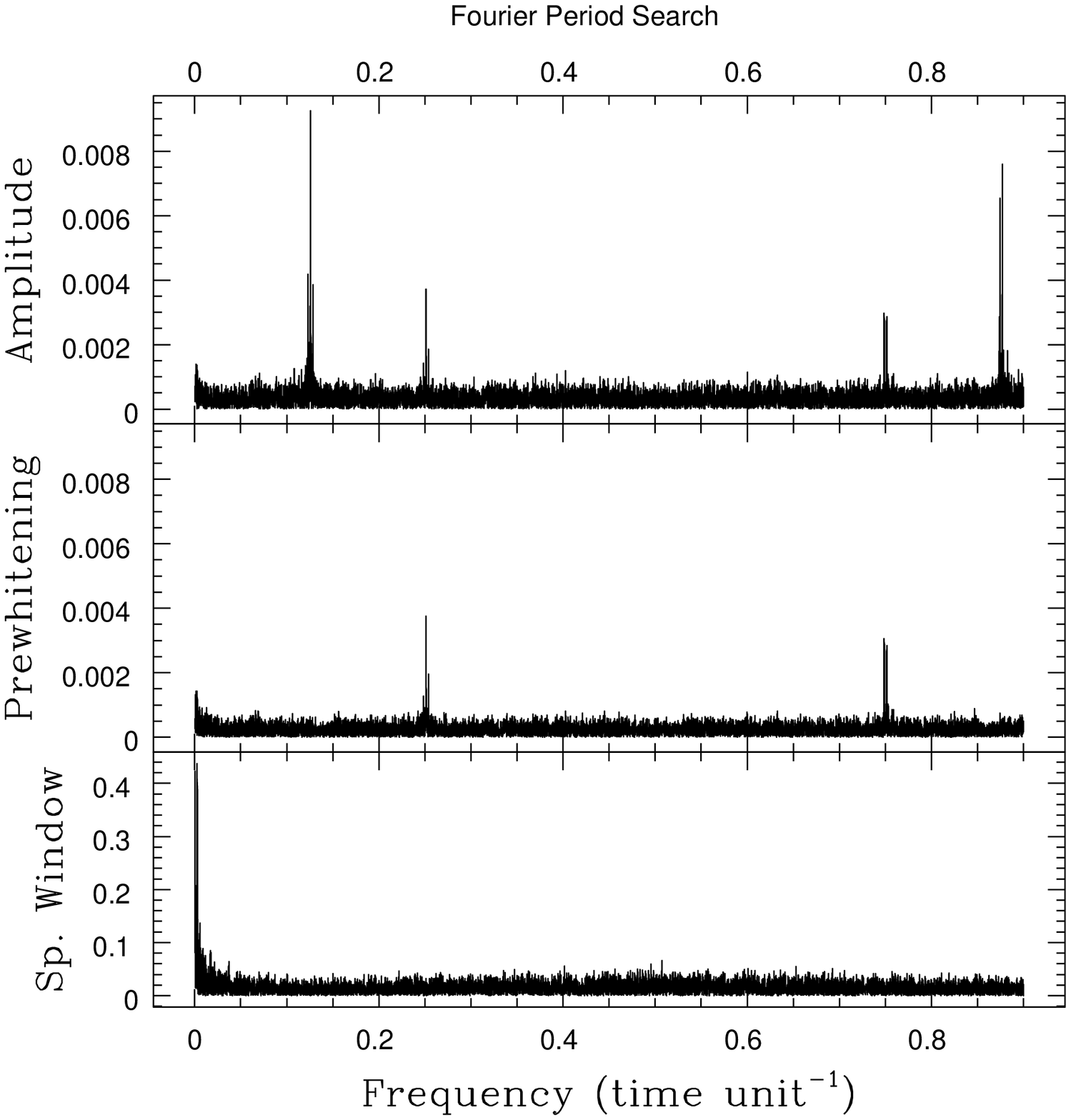}
\includegraphics[width=6cm]{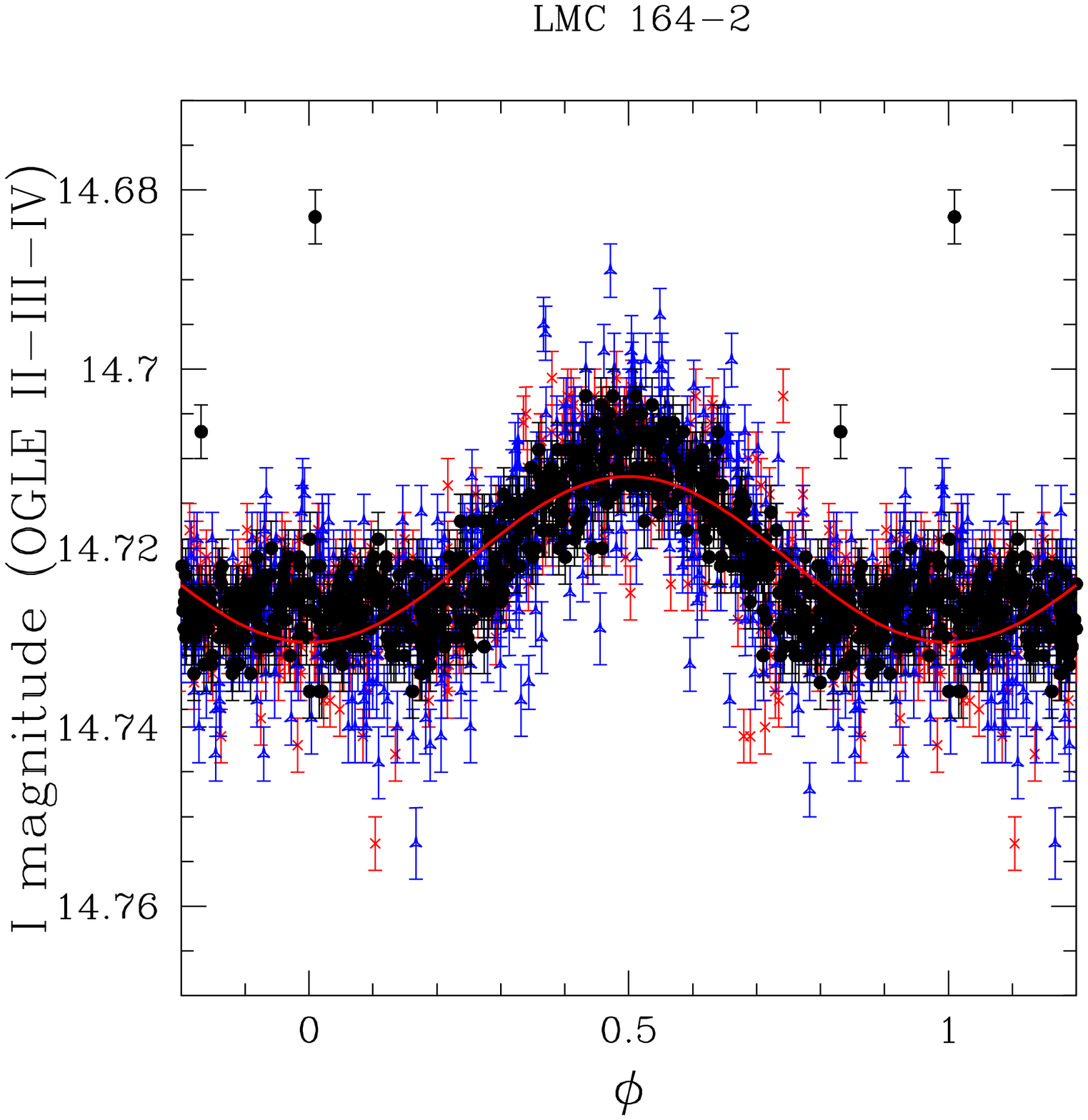}
\caption{Same as Fig. \ref{2dfsbis} for LMC164-2 (see Table \ref{list} for ephemeris). We note in the left panel the presence of peaks at frequencies corresponding to multiples of the fundamental period for all methods except Fourier, and in the middle panel the best-fit signal at 0.126\,d$^{-1}$, its second harmonic at 0.25\,d$^{-1}$, and their daily aliases at 0.87\,d$^{-1}$ and 0.75\,d$^{-1}$, respectively. }
\label{lmc1642}
\end{figure*}

A $\chi^2$ test shows that LMC164-2 is significantly variable in both ASAS and OGLE datasets. All period searches on OGLE data reveal a very clear peak at $P=7.96$\,d: values range from 7.9598d for variance methods to 7.9612\,d for Renson's method, which is fully consistent with the 0.0010\,d $1\sigma$ error (see Table \ref{list} and left panel of Fig. \ref{lmc1642}). Moreover, this period yields coherent variations with phase and, when eliminated from the data, leaves a flat periodogram (Fig. \ref{lmc1642}, middle and right panels). The EROS-2 data in blue band (where most data points are available) of LMC164-2 have the lowest errors amongst EROS-2 data (always$<$0.045\,mag). Formally, they do not deviate significantly from a constant, but period search methods readily confirm the signal found in OGLE data, with a period of 7.959$\pm$0.003\,d and an amplitude in this band of 0.0089$\pm$0.0016\,mag. Periodograms and phased variations are very similar to those shown in Fig. \ref{lmc1642}. This signal is thus undoubtly real, though it remains buried in ASAS noise. 

\subsection{SMC159-2}
\begin{figure*}
\includegraphics[width=6cm]{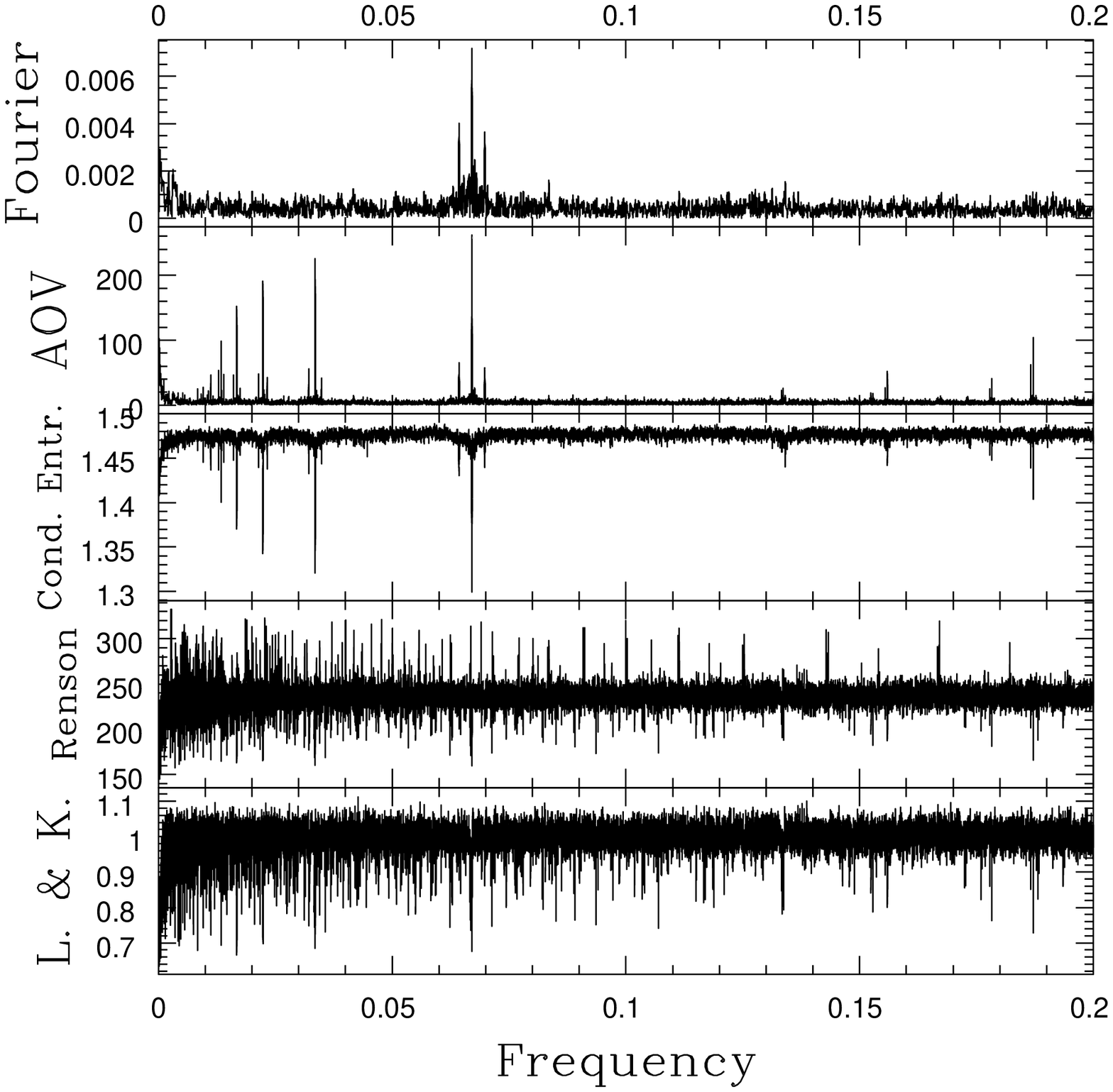}
\includegraphics[width=6cm]{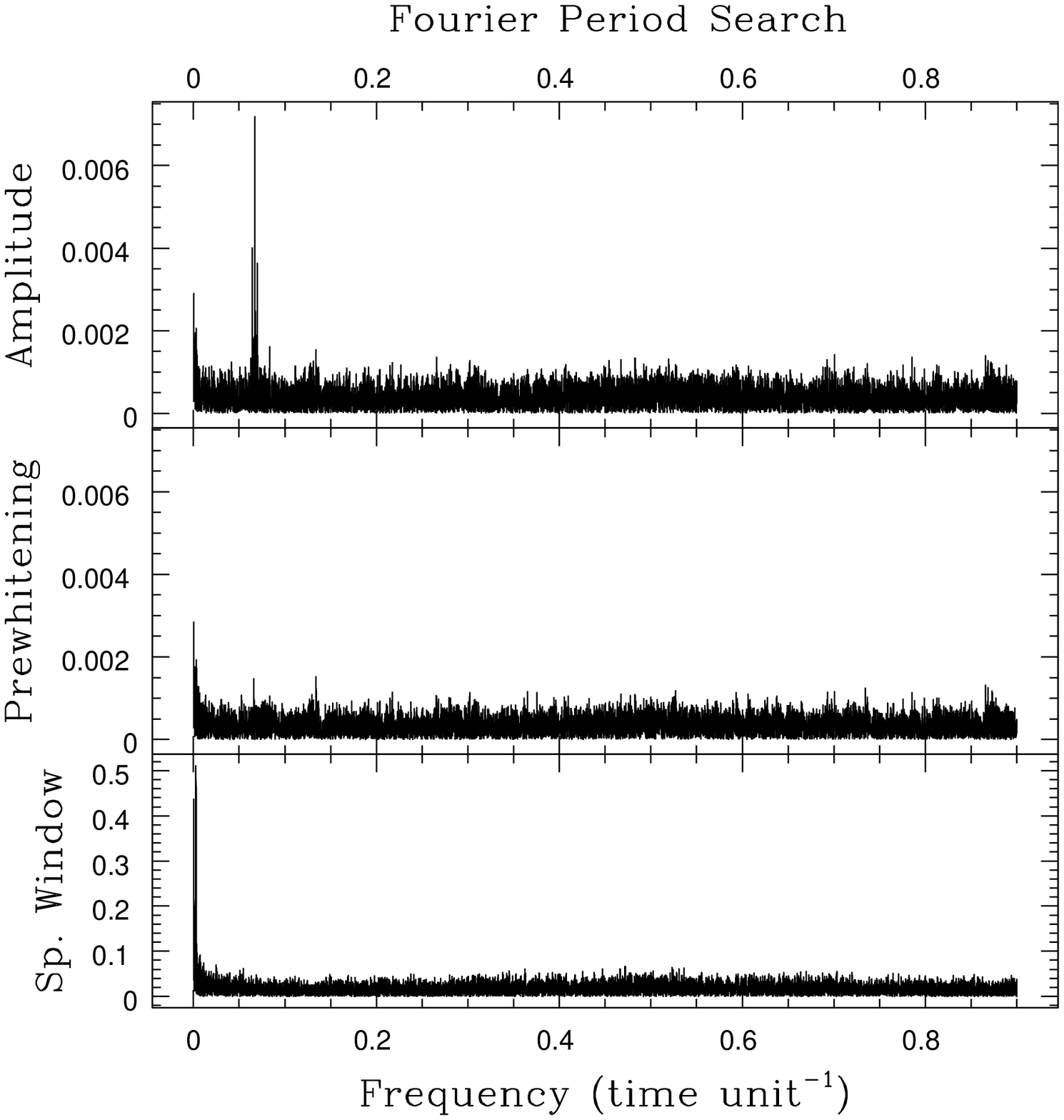}
\includegraphics[width=6cm]{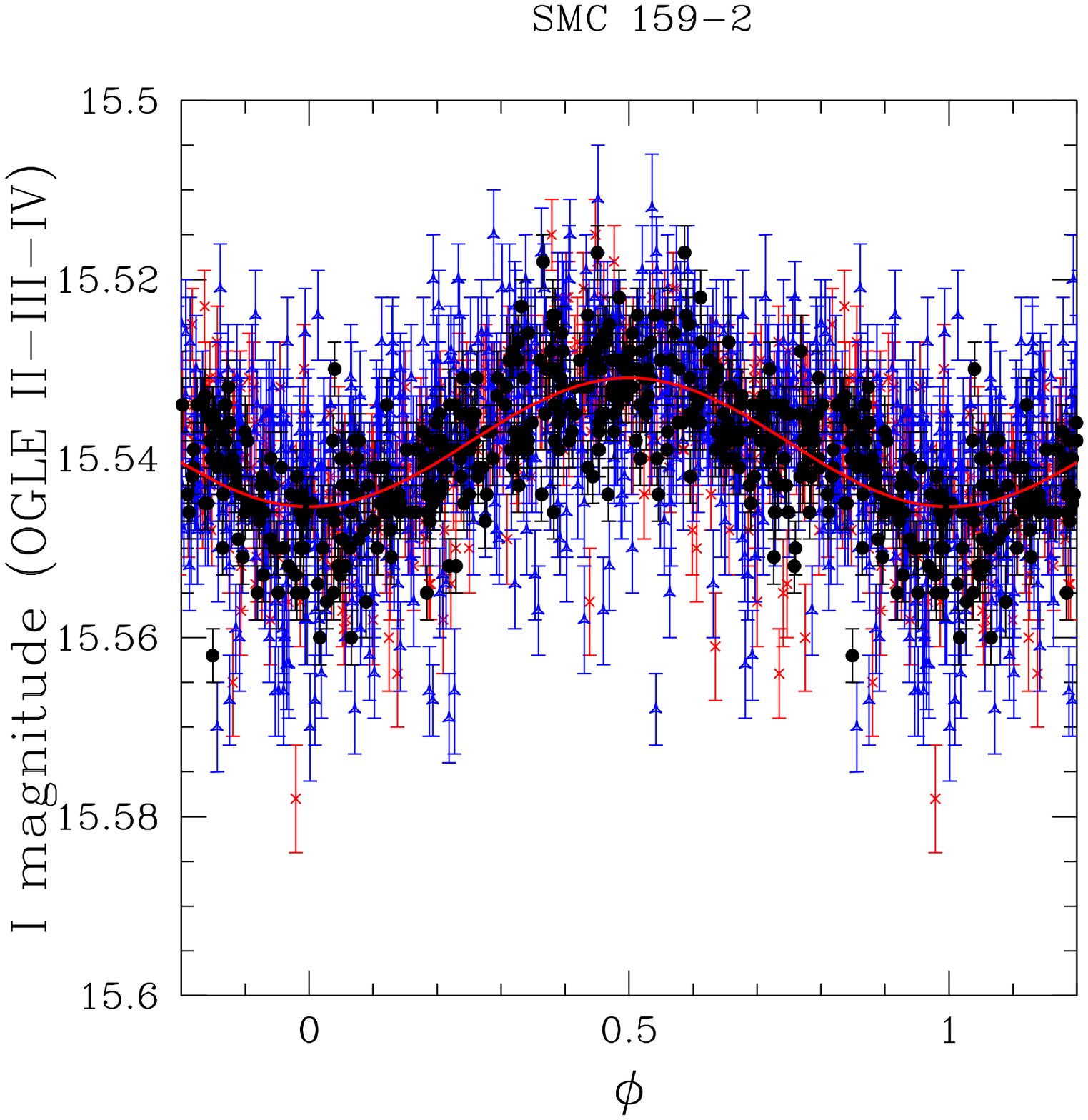}
\caption{Same as Fig. \ref{2dfsbis} for SMC159-2 (see Table \ref{list} for ephemeris). }
\label{smc1592}
\end{figure*}

SMC159-2 appears only in the OGLE database and the OGLE data appear significantly variable, as shown by a $\chi^2$ test. Period searches on OGLE data reveal a very clear peak at $P=14.91$\,d : values range from 14.912\,d for variance methods to 14.915\,d for the conditional entropy method, a difference which is well within the 0.004\,d $1\sigma$ error (see Table \ref{list} and Fig. \ref{smc1592}). This period yields coherent phased variations and, when eliminated from the data, leaves a flat periodogram (Fig. \ref{smc1592}). Again, this leaves little doubt about the reality of the signal. 

\begin{figure*}
\includegraphics[width=15cm]{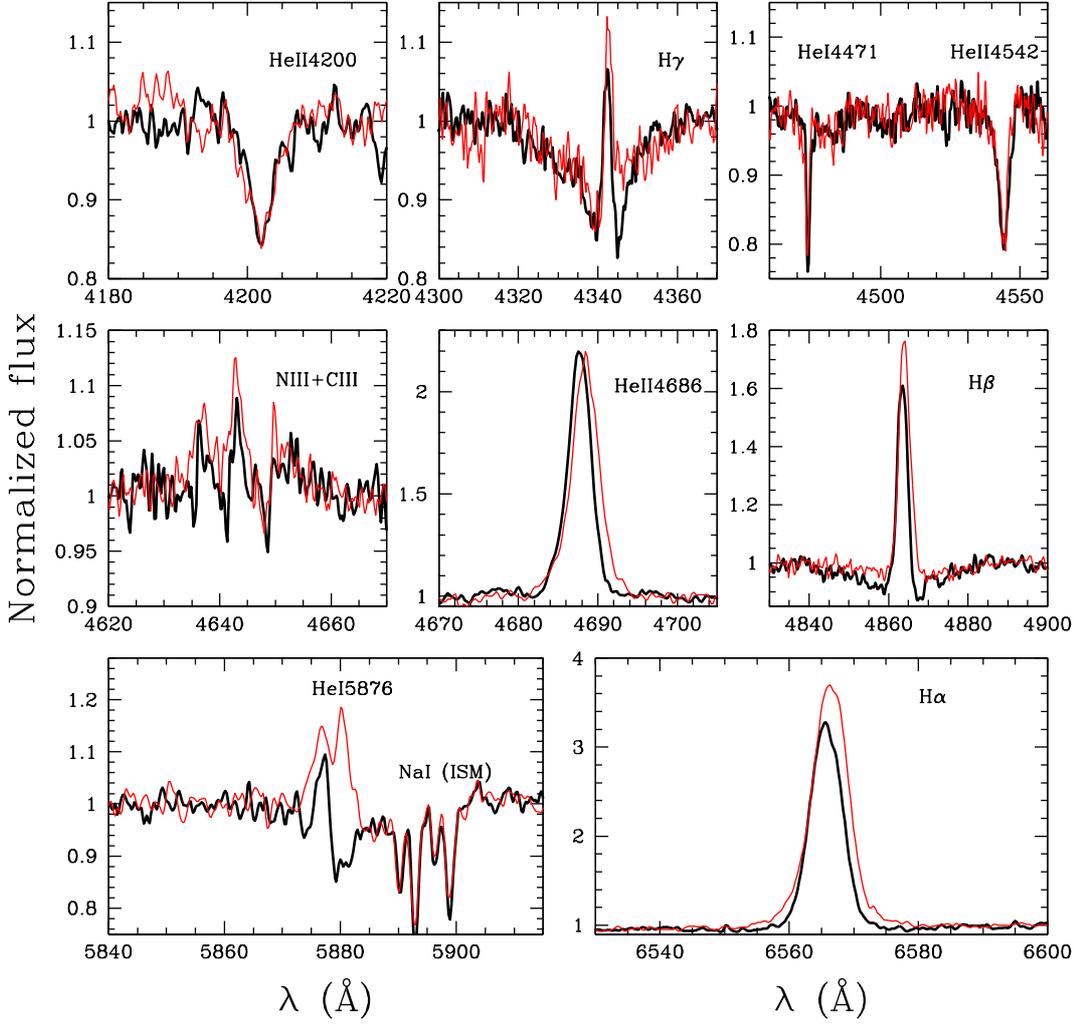}
\caption{Spectra of SMC159-2 recorded in 2013 ($\phi$=0.96, thick black line) and 2014 ($\phi$=0.60, thinner red line) in selected spectral windows:
clear changes in Balmer and \hei\ lines can be seen, with stronger emissions in 2014, while photospheric \heii\ absorptions and interstellar lines arising in both the Galaxy and the SMC (e.g. Na D lines near 5900\,\AA) remain constant.  }
\label{specsmc}
\end{figure*}

\begin{figure}
\includegraphics[width=8cm]{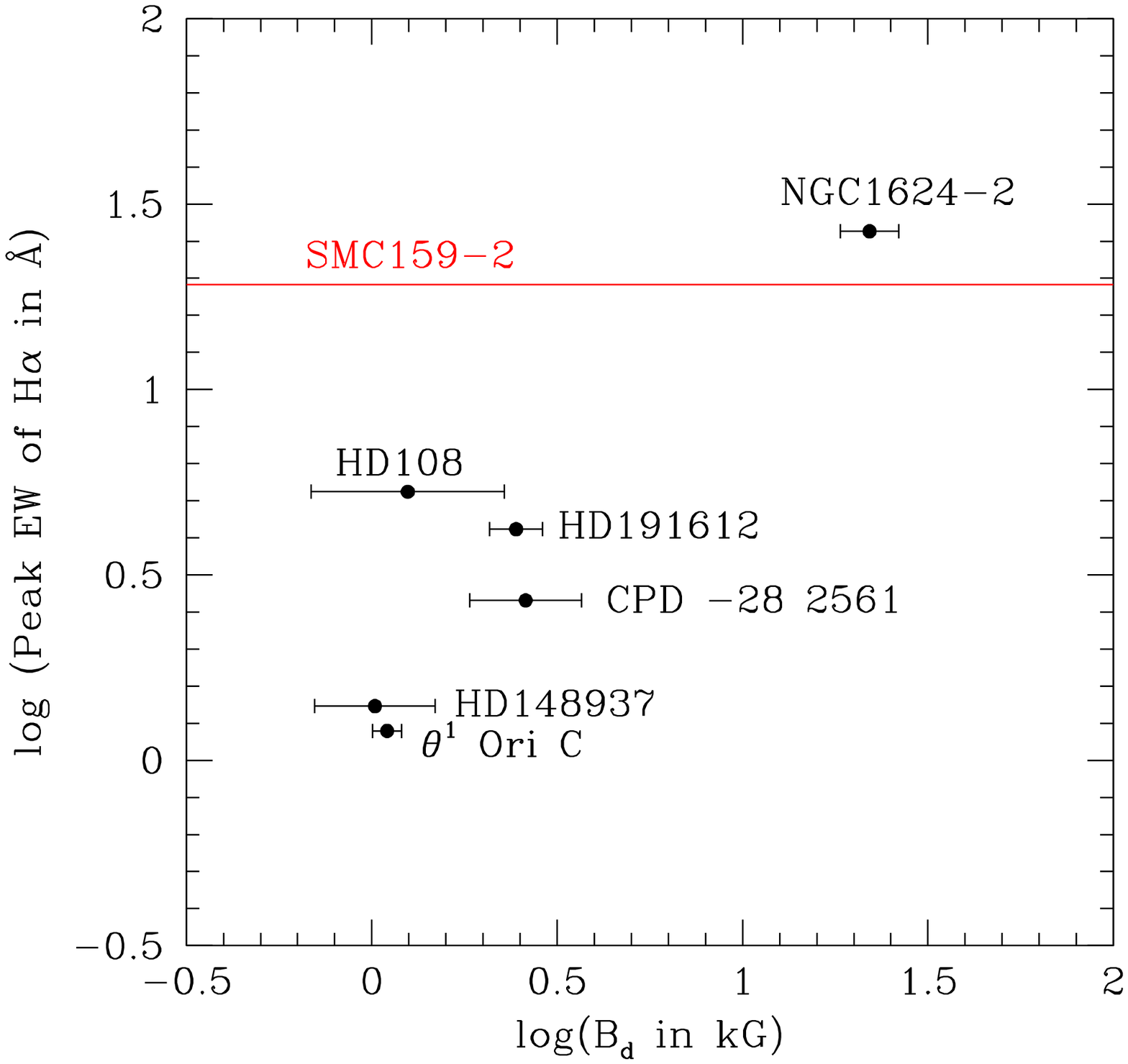}
\caption{Comparison of the peak \ha\ EW and the dipolar field strength of magnetic O stars (using unpublished spectropolarimetry obtained by GAW for NGC1624-2). The horizontal red line indicate the peak \ha\ EW of SMC159-2.  }
\label{EWB}
\end{figure}

Two spectra of SMC159-2 are available, one taken in 2013 (on $HJD=$2\,456\,640.551) and one in 2014 (on $HJD=$2\,456\,903.691). Using the best-fit ephemeris derived from Fourier algorithm (Table \ref{list}), we found that the 2013 and 2014 spectra of SMC159-2 were acquired at two very different phases, 0.96 and 0.60, respectively. These spectra display many features that are typical of Galactic Of?p stars (Fig. \ref{specsmc}). The \heii\ absorption lines remain essentially constant, indicative of a predominantly photospheric origin, and they are broader than \hei\ absorption lines, as typically observed in Of?p spectra. We note that the apparent spectral type does not change much, the \heii\,$\lambda$\,4541/\hei\,$\lambda$\,4471 ratio remaining consistent with a spectral type of O6. In addition, the emission lines appear relatively narrow (e.g. FWHM of $\sim$280\,km\,s$^{-1}$ for \ha), i.e. much narrower that typical wind \ha\ emission of O-stars, as is usually observed in Of?p stars. In parallel, many variations typical of Of?p stars are also observed. The Balmer lines and \heii\,$\lambda$\,4686 emission lines are stronger in 2014, although with one exception (HD\,148937) the differences are generally smaller than in the spectra of Galactic Of?p stars. These lines appear slightly redshifted at maximum, as is seen e.g. in CPD$-28^{\circ}$2561 \citep{wad15}. In addition, the \niii\ and \ciii\ emission lines near 4630--4650\,\AA\ are stronger in 2014, the \ciii\ emissions nearly disappearing in 2013. The most extreme variation occurs in the metastable \hei\,$\lambda$\,5876 line, which changes from an apparent inverse P~Cygni profile at minimum, to a much stronger double emission line at maximum, the second peak appearing longward of the single minimum emission. This could be related to the redshifts observed for other emission lines. If this star is confirmed to be magnetic, then the physical origins of these diverse behaviours are linked to the geometry of the rotating magnetospheric structure. A full spectroscopic monitoring will thus be needed to derive the stellar properties, as was done e.g. for HD\,191612 \citep{wad11}. 

In the Galaxy, Of?p stars brighten when their emission increases \citep{wal04,bar07}. The phase of the 2013 spectrum of SMC159-2 ($\phi$=0.96) corresponds to a brightness minimum, while the phase of the 2014 spectrum ($\phi$=0.60) indicates that it was taken close to brightness maximum. Furthermore, a direct comparison is possible as the last photometric data were acquired close in time to the 2013 spectrum: the I-magnitude was 15.544$\pm$0.003 on $HJD$=2\,456\,639.571, i.e. the star was indeed faint at the time. There is thus a direct correlation between the strength of the emission lines and the brightness of the star: SMC159-2 behaves in a manner similar to the Galactic Of?p stars.

Equivalent widths (EWs) of \ha\ and \heii\,$\lambda$\,4686 were evaluated by integrating the spectra in the 6550--6600\,\AA\ and 4680--4697\,\AA\ intervals, respectively, as well as by fitting gaussians to the emission peaks. Derived EWs amount to 13.7$\pm$0.5\,\AA\ and 4.3$\pm$0.2\,\AA\ in 2013 and to 19.2$\pm$0.2\,\AA\ and 5.0$\pm$0.2\,\AA\ in 2014 for \ha\ and \heii\,$\lambda$\,4686, respectively. It must be noted that there is no nebulosity near SMC159-2, hence the recorded Balmer and \hei\ lines are uncontaminated: the emissions and their variability are truly linked to the star. We also note that the emission in \ha\ is very strong, suggesting a large quantity of emitting material. Amongst magnetic O stars, the peak EW of \ha\ emission generally amounts to a few \AA ngstr\"oms \citep[e.g.][]{how07}, except for the exceptional case of the highly magnetized NGC1624-2 which presents a peak EW of about 27\,\AA\ \citep{wad12dash}. Figure \ref{EWB} shows the peak EW of the \ha\ line for $\theta^1$\,Ori\,C and the five Galactic Of?p stars as a function of their surface magnetic field strengths $B_d$. This figure demonstrates the expected increase in emission with the field strength, and hence with the magnetospheric volume. For slow rotators such as Of?p stars, this magnetospheric volume is determined by the Alfv\'en radius: it is thus a function of both the magnetic field strength and wind momentum, but the latter was assumed to be approximately uniform for all stars in Fig. \ref{EWB}. A forthcoming paper (Wade et al, in prep) will examine in more detail the theoretical expectations for the MC Of?p stars.  Based on this figure and the outstanding strength of its \ha\ emission, we can predict that the magnetic field of SMC159-2 is very strong, second only to that of NGC1624-2 and likely in the range 5--15\,kG.

\section{Conclusions}
The Of?p category comprises magnetic massive stars, where an oblique magnetic field provokes rotationally-modulated photometric and spectral variations in the optical, UV and X-ray domains. Up to now, such objects were only studied in detail in the Galaxy, but five candidate extragalactic Of?p stars have been proposed. No spectropolarimetry of these objects is yet available, and further spectroscopic data, though somewhat scarce, will be soon presented by Walborn et al. (in preparation). 

In this paper, we analysed the photometry of these five objects. Three of them behave exactly like magnetic Of?p stars in the Galaxy: the photometric amplitudes, the roughly sinusoidal form of the variations, and the strict and stable periodicity with periods from 8d to 1370d are similar to what is found in Galactic cases. These stars thus have a high probability of being magnetic, though it is difficult to derive the system's geometry and magnetic field strength with the photometric data alone. Moreover, simultaneous spectral variations observed in SMC159-2 are also reminiscent of the behaviour of Galactic Of?p stars. It can be noted that the \ha\ line of SMC159-2 is remarkably strong. This leads us to predict that its magnetosphere is very large and that its magnetic field is especially strong, likely only second to that of the outstanding star NGC 1624-2. 

For the remaining two Of?p candidates, significant photometric variability is observed, but strict periodicity cannot be ascertained. Their unexplained photometric variations may well be a consequence of multiple variability origins (e.g. rotation in combination with binarity or pulsation). Therefore, while these stars demonstrate the spectral peculiarities expected of hot magnetic Of?p stars, we are not yet able to confirm photometric variability consistent with stable, long-term rotational modulation. 

This study paves the way for further observations, opening the door to the study of magnetic field and stellar wind interactions in extragalactic environment at lower metallicities.

\begin{acknowledgements}
We warmly thank Phil Massey for having uncovered three of the Of?p candidates and for sharing data as well as observing time with the authors. YN thanks E. Gosset for useful discussion and acknowledges support from  the Fonds National de la Recherche Scientifique (Belgium), the PRODEX XMM contracts, and the ARC (Concerted Research Action) grant financed by the Federation Wallonia-Brussels. GAW acknowledge Discovery Grant support from the Natural Science and Engineering Research Council of Canada (NSERC). M.K. Szymanski acknowledges funding of the OGLE project from the European Research Council under the European Community's Seventh Framework Programme (FP7/2007-2013) / ERC grant agreement no. 246678. This work also made use of EROS-2 data, which were kindly provided by the EROS collaboration, thanks to J.B. Marquette. The EROS (Exp\'erience pour la Recherche d'Objets Sombres) project was funded by the CEA and the IN2P3 and INSU CNRS institutes. ADS and CDS were used for preparing this document. 
\end{acknowledgements}

\end{document}